\documentclass[11pt]{article}
\usepackage{latexsym}
\begin{document}
\begin{center}{\Large {\bf Return Probability of a Random walker
 in continuum with uniformly distributed jump-length}}\end{center}

\vskip 1cm

\begin{center}{\it  Ajanta Bhowal Acharyya}\\
{\it Department of Physics, Lady Brabourne College}\\
{\it P-1/2 Suhrawardy Avenue, Calcutta - 700017, India}\\
{E-mail:ajanta.bhowal@gmail.com}\end{center}

\noindent 
{\bf Abstract:} We are studying the motion of a random walker in two and three 
dimensional continuum with uniformly distributed jump-length. 
{\it This is different from conventional Levy flight}.
 In 2D and 3D continuum, a random walker can move in any
 direction, with equal probability  and with any step-length($l$) varying 
randomly between 0 and 1 with equal probability.  A random walker, who starts its
 journey from a particular point (taken as origin), walks through the region,
 centered around the origin.  Here, in this paper, we studied the probability 
distribution of end-to-end distances and the probability of return for the 
first time within a
 circular zone of finite radius, centered about the initial point, of a random
 walker.  We also studied the same  when a random walker moves with constant
 step-length, unity.
The probability distribution of the time of first return (within a specified
zone) is also studied.

\newpage

\noindent {\bf I. Introduction:}

The diffusion of a particle through the fluid is an age old but challenging
problem\cite{binder}. The statistical physicists try to understand this by 
modelling and computer simulation of random walk \cite{simple}.
In simple random walk model, every possible new step has the same probability.
The random walk on  higher dimension  is studied exhaustively considering the
motion of the random walker takes place on  regular lattice.
In d-dimension, the mean square displacement, $<d^2(t)>$, 
of a walk  after '$t$' time step varies with time ($t$) according to
the relation $<d^2(t)>\sim t$.

The random walk (particularly, self avoiding walk)\cite{poly} is used to 
study the statistical properties of the configurations of flexible
macromolecules in a  good solvent.  The directed self avoiding walk has
been studied extensively to analyse the process of polymerisation. 
In two dimensions, the size of the polymer scales with the number of
monomer in a power law fashion\cite{mono}.
Random walk problem has also been studied thoroughly on Bethe 
lattice\cite{bethe}. The biased random walk has also been studied recently
\cite{bias} in biological system.
Boyer and Solis-Salas recently studied and solved exactly the non-markovain
 random
walk which helps to understand the animal and human mobility\cite{Denis}.

It is to be mentioned that in  all the cases of random walk, the model was
 built on a regular lattice. Precisely,
 the walker moves with  unit step-length and along  specified finite number of
 directions. But in reality,  a particle
generally moves in any direction with any value (distributed and bounded) of 
step-length. It is quite natural expectation that the random walk in continuum
would be able to be more analogous to the reality. 

In this particular article, the random walk on two and three dimensional
 continuum is studied by Monte Carlo simulation. Here we consider both the
 situation:(i) the length of step is unity and (ii) step-length  is taken to
 be distributed uniformly and bounded between 0 and 1. 
In case of random walk on 2D/3D continuum, the probability of return to a
circular/spherical  zone, centered about the walkers' initial  point,
 is calculated for different values of the radius of the zone. 
The probability distribution of the time of first return (within a specified
zone) is also studied.

The manuscript is organised as follows: in the next section
(Section-II) the simulation scheme and the numerical results are reported,
the paper ends with concluding remarks in the section-III.

\newpage

\noindent {\bf II. Simulation and Results}

For simple random walk in 2D regular lattice, the random walker can move in any
one of the four (up, down, right, left) direction with equal probability.
Whereas in case of random walk in continuum, the random walker  can move in
any direction, with equal probability  and with step-length ($l$).
Suppose $x(t)$, $y(t)$ is the position of the random walker at time $t$, then
in the next time step, the position of the random walker will be as follows:
$$x(t+1)=x(t)+l \cos(\theta) ~~~~~~{\rm and}$$ 
$$y(t+1)=y(t)+l \sin(\theta)~~~~~~~~~~~~~~~$$ 
where $l$ is the step-length and $\theta$ is a  variable which determines
the direction of the walk.
Here, $\theta$  can have any value between 0 and $2\pi$
with equal probability.  We studied here both the cases, one with
constant step-length ($l=1$) and the other with random step-length $l$, which
can vary between  0 and 1 with equal probability.
 It is to be mentioned here that the  step-length equals to zero
means that the random walker does not move. Thus, in this model, we consider the
situation that the random walker has a choice not to move.
This generalisation of the problem of random walk  makes it an interesting
problem of statistical computational physics.  

As we know that the random walk in continuum is nothing but the solution of 
diffusion equation.  Still for the verification, 
we have calculated, how does the mean square displacement,
 $<d^2(t)>$, of a walk vary with time in case of 2D and 3D continuum.

We calculate the mean square displacement of a random walk, in 2D continuum,
by averaging the square of the displacement over $N_s$ such random walker.
Fig.1 plots the variation of $<d^2(t)>$, mean square distance of the final 
position ($N_s =1000$ ) with time for both type of 
step-length of the random walk.
The linear nature of the graph verifies the diffusive behaviour, 
$<d^2(t)> \sim t$.
From Fig.1, it is observed that  the increase of the displacement with time 
is
more in case of uniform step-length compared to the case of randomly varying 
step-length, which is quite obvious.
It is also obvious that the displacement of different  random walkers  after
a particular  time, $t$, are different. Now let us  see how these
 displacements are distributed.
Fig.2a gives the normalised distribution of the displacement ,$d(t)$,
of random walk in continuum having step-length unity($l=1$), after a 
finite time, $t=100000$ and $t=225000$. 
The distribution is calculated for $N_s$ random walkers and here $N_s=t$.
Fig.2b gives the normalised distribution of the displacement, $d(t)$,
 for randomly varying step-length ($l$), after a finite time
 $t=100000$ and $t=225000$. 
Here also the distribution is calculated for $N_s$ random walkers ($N_s=t$).
In Fig.3 we have plotted the rescaled distribution of displacement, 
distributions of ${d(t)\over {\sqrt t}}$
for the above  mentioned values of time $t=100000$ and $225000$ for unit
 step-length and randomly varying step-length on the same graph.
Fig.3  shows that the mode of the distribution as well as the width of the 
distribution of displacement in case of randomly varying step-length 
is less compared to  the case of random walk with uniform step-length.

It is well known that a simple random walker on 2D regular lattice must
return to the initial point after a finite time, though this time of returning
is different for different random walkers.
In 3D regular lattice only a fraction of random walkers will return to the 
initial point after a finite time. Here,   
we are interested to calculate the  probability of return for the first time 
 in case of random walk in continuum.
Since in case of random walk in continuum, there are infinitely many paths,
it is really impossible to check whether a random walker come back exactly 
at the initial point.
Thus, in this model of continuum random walk,  we have calculated the 
probability of return within a circular zone, centered around the initial point
 instead of exactly at the initial point in  case of random walk on regular lattice.
At this point it is also to be mentioned that for the calculation of returning 
probability in the case of randomly varying step-length,  we assume that each 
of the random walkers move their first step  with step-length, $l=r_z+\zeta$
, where $\zeta$  is a random number uniformly distributed between 0 and 1.
This confirms that after the first step, all the random walkers are outside
the initial central  zone. 
Thus the probability of return, $P_r$, is the  fraction of random walkers 
who will enter to this  circular zone of radius $r_z$,
 centered around the initial 
point. 

In this paper, we have also studied the various kinds of behaviour
 of  probability of return for the first time  for random walk in 2D continuum
having unit step-length. 
Here, we have studied how does  the  probability of return for the first time 
to the initial zone vary with time for a particular radius of the  circle
 for both type of step-length.
Fig.4 shows the variation of return probability with time for a
particular value of of the radius of the circle ($r_z=.5$) for 
both type of variation of step-length.
It is quite obvious that the return probability, $P_r$,  increases as the
 radius of the central zone increases. 
For the calculation of the probability of return we consider  all those random
walkers who entered the central zone after first step. Among these first 
returned walkers, there are walkers, who entered the central zone just immediately after the first step, i.e, at the time, $t=2$. We first calculated the
distribution of the first returned time. We also calculated the fraction
of walkers',  who immediately returned to the central zone. It is observed
that for 2D regular lattice,  this value is .24 , for random walk in continuum
with unit step-length, the value is .16 and that for random step-length .094.
This shows that as the no of path  increases, the fraction of immediate 
return decreases.

In Fig.5, we plotted the variation of probability of return for the first time,
$P_R$, with the radius of the central zone ($r_z$) for a particular time. 
Here the return probability is calculated for time, $t= 10000$ and $N_s=100000$. 
We also calculated the probability distribution of first return time
(${\tau}_R$) in 2D continuum. Here, we also calculated the probability of
first return to the initial point in 2D regular lattice. 
Fig.6  gives the plot (in logscale) of the distribution of first return time
(${\tau }_R$) to the central zone of radius $r_z=0.5$.
The linearity of the plot in logscale reveals the scale invariant nature
of the probability distribution of first 
return  time, i.e., $P({\tau}_R)\sim {\tau}_R^{-\alpha}$ with $\alpha =1.2$.

We have also studied the random walk in 3D continuum for both type of 
step-length as in the case 2D continuum.
In case of random walk in  3D continuum, the random walker  can move in
any direction, with equal probability  and with step-length ($l$).
Suppose $x(t)$, $y(t)$, $z(t)$ is the position of the random walker at time $t$, then
 the position of the random walker at next time  will be updated as follows:
$$x(t+1)=x(t)+l \sin(\theta) \cos(\phi)$$ 
$$y(t+1)=y(t)+l \sin(\theta) \sin(\phi)$$ 
$$z(t+1)=z(t)+l \cos(\theta) ~~~~~~ {\rm and}$$ 
where $l$ is the step-length and $\theta$, $\phi$ are   variables  which
 determines the direction of the walk.
Here, $\theta$  can have any value between 0 and $\pi$
and  $\phi$  can have any value between 0 and $2\pi$
with equal probability.  We studied here both the case, one with
constant step-length ($l=1$) and the other with random step-length $l$, which
can vary between  0 and 1 with equal probability.

Here, in the case of random walk in 3D continuum,
 also we have also calculated the probability distribution  of
 end-to- end distances  or displacements after time $t=100000$.
It is calculated for $N_s$ random walkers, ($N_s=t$).
In Fig.7 we have plotted the distribution of the rescaled displacement, i.e. distribution
 of ${d(t)\over {\sqrt t}}$   upto $t=100000$ for  both type of step-length
 in case of random walk on  3D continuum as well as for 3D regular lattice
 in the same graph.
Like, the case of 2D random walk, it also  shows that the mode of the
 distribution as well as the width of the distribution of displacement
becomes less as the step-length  changes from uniform to a randomly varying one.
It is also interesting, that the introduction of various possible directions
of random walk in the continuum does not change the distribution, whereas
 the randomness in the step-length  changes it remarkably.

In this article, we calculate  the probability of return for the
first time to the central zone. To calculate this we consider a  spherical
zone of radius $r_z$, centered around the initial point of the random walker
instead of a circular zone in case of 2D continuum.
In Fig.8, we plotted the variation of probability of return for the first time,
$P_R$, with the radius of the central zone ($r_z$) for a particular time. 
Here the probability of return for the first time is calculated for 
time, $t= 10000$ and $N_s=100000$. 

In Fig.9  we  plotted (in logscale), the distribution of first return time
(${\tau }_R$) to the central zone of radius $r_z=0.5$.
The linearity of the plot in logscale reveals the scale invariant nature
of the probability distribution of first 
return  time, i.e., $P({\tau}_R)\sim {\tau}_R^{-\alpha}$ with $\alpha =1.32$.

\vskip 1cm

\noindent {\bf III. Concluding Remarks}

In this paper, the random walk on two and three dimensional continuum
is studied by Monte Carlo simulation. Here the length of jump is considered
to be random and distributed uniformly and bounded between 0 and 1.  Thus the
distance between two successive positions of a random walker varies between 
0 and 1.  The following observations are made: 
The mean square displacement is studied as a function of time.
As expected,  this shows the diffusive behaviour like the simple random walk on lattice.
The rescaled distribution of the displacement (of the final position)
is also studied. This shows a nonmonotonic variation having a most probable
value of the displacement. 
 As step-length of the random  walker becomes unity instead of randomly varying between 0 and 1,  the position of the 
most probable value increases and the width of the distribution also
increases. The probability of return  for the first time to a zone
of a circle centered about the starting point is calculated for  different
values of the radius of the zone. For a particular value of radius of the 
central zone, the return probability initialy increases with time and then
after a long time it becomes almost constant.  Here, it is also observed that
 as the radius of the zone increases the probability of return for the 
first time  increases. 
In case of random walk on planar continuum the probability of return approaches
 unity. Whereas the value of return probability in 3D continuum  approaches 
  a value, which differs from 3D regular lattice (.34).
The probability distribution of the time of first return (${\tau}_R$) of
 a walker shows scale invariance nature.

There are several important and relevant aspects can be studied. For
example, to study the fractal dimension (if any), first passage time 
etc.

\newpage
% GNUPLOT: LaTeX picture FIG-1 <d^2> vs t in 2D
\setlength{\unitlength}{0.240900pt}
\ifx\plotpoint\undefined\newsavebox{\plotpoint}\fi
\sbox{\plotpoint}{\rule[-0.200pt]{0.400pt}{0.400pt}}%


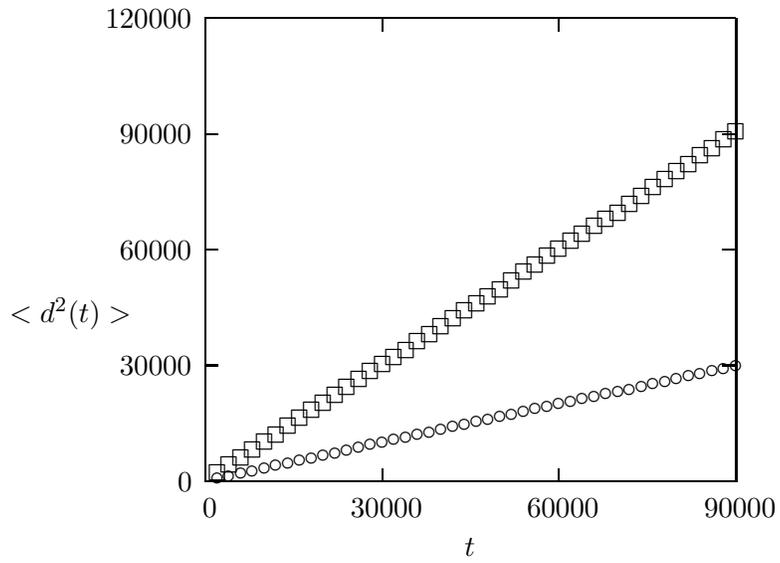
\noindent {\bf Fig-1.} The mean square displacement $<d^2(t)>$ plotted against
time ($t$)  for random walk in 2D continuum. ($\Box$) represents the uniform 
step-length and ($\circ$) represents the random step-length. 

\newpage

% GNUPLOT: LaTeX picture FIG-2a P(d) uniform 2D
\setlength{\unitlength}{0.240900pt}
\ifx\plotpoint\undefined\newsavebox{\plotpoint}\fi
\sbox{\plotpoint}{\rule[-0.200pt]{0.400pt}{0.400pt}}%
%

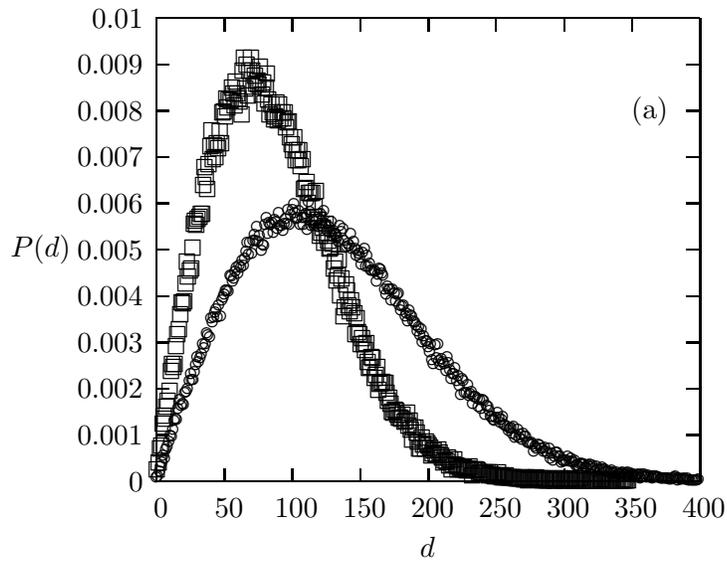
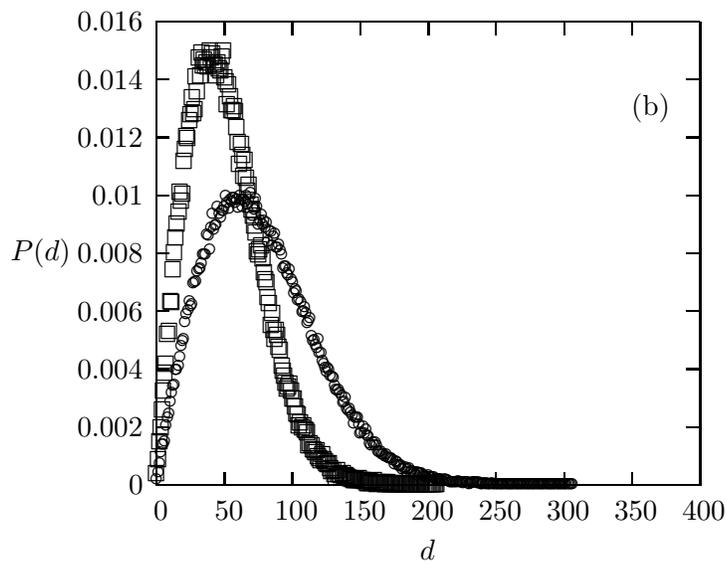
\noindent {\bf Fig-2.} The distribution of displacement in 2D continuum.
(a) for uniform length-step ($\Box$) t=100000, ($\circ$) t= 225000 
and (b) for random length-step ($\Box$) t = 100000, ($\circ$) t=225000.

\newpage
% GNUPLOT: LaTeX picture FIG-3: p(d\sqrt(t)) vs d\sqrt(t) for both
\setlength{\unitlength}{0.240900pt}
\ifx\plotpoint\undefined\newsavebox{\plotpoint}\fi
\sbox{\plotpoint}{\rule[-0.200pt]{0.400pt}{0.400pt}}%
%

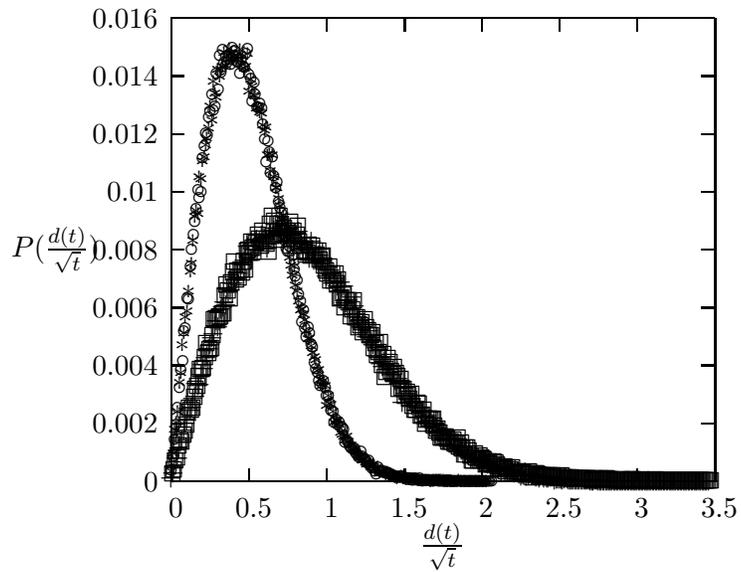
\noindent {\bf Fig-3.} The probability distribution of rescaled displacement in 2D continuum.
For uniform length-step ($\Box$) t=100000, ($+$) t= 225000 
and for random length-step ($\circ$) t = 100000, ($\ast$) t=225000.
\newpage
% GNUPLOT: LaTeX picture FIG-4 Prob of return vs time in 2D
\setlength{\unitlength}{0.240900pt}
\ifx\plotpoint\undefined\newsavebox{\plotpoint}\fi
\sbox{\plotpoint}{\rule[-0.200pt]{0.400pt}{0.400pt}}%
%

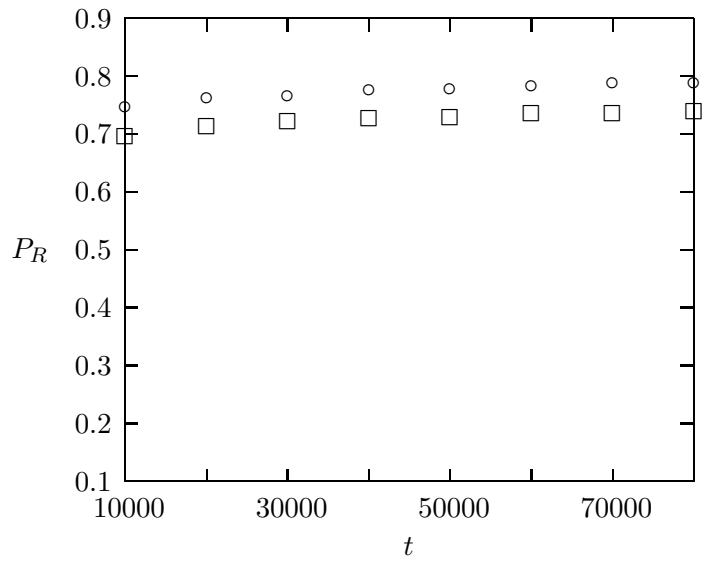
\noindent {\bf Fig-4.} The time variation of probability of return for the first time  to the central zone of radius $r_z=.5$ in 2D continuum.
For uniform length-step ($\Box$) and for random length-step ($\circ$).
\newpage
% GNUPLOT: LaTeX picture FIG-5 Variation of return prob with rlim
\setlength{\unitlength}{0.240900pt}
\ifx\plotpoint\undefined\newsavebox{\plotpoint}\fi
\sbox{\plotpoint}{\rule[-0.200pt]{0.400pt}{0.400pt}}%
%

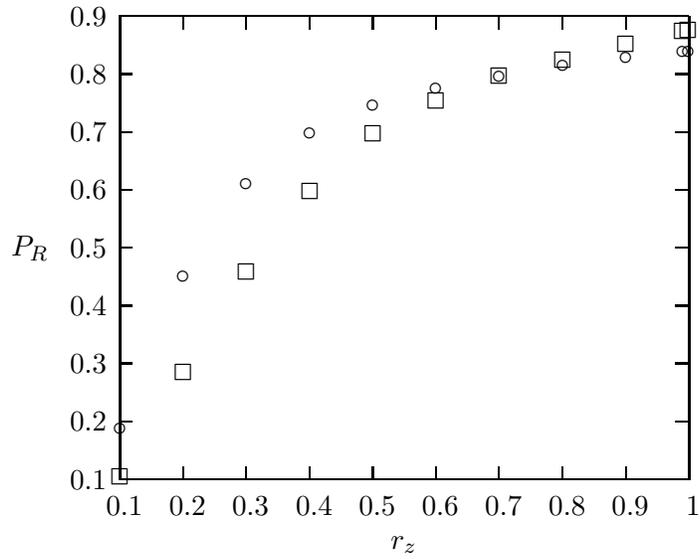
\noindent {\bf Fig-5.} The variation of probability of return for the first time  to the central zone with  radius of the zone, $r_z=.5$ in 2D continuum for time, $t=10000$.
For uniform length-step ($\Box$) and for random length-step ($\circ$).

\newpage

% GNUPLOT: LaTeX picture FIG-6 Log-log plot of ditribution of first return time
\setlength{\unitlength}{0.240900pt}
\ifx\plotpoint\undefined\newsavebox{\plotpoint}\fi
\sbox{\plotpoint}{\rule[-0.200pt]{0.400pt}{0.400pt}}%
%

\noindent {\bf Fig-6.} Log-log plot of the  probability distribution  of first 
return  time  to the central zone for a particular  radius of the zone, $r_z=.5$ in 2D continuum  and for regular lattice for time, $t=10000$ and $N_s=100000$.
For uniform length-step ($\Box$), random length-step ($\circ$) and for regular
2D lattice ($\ast$). The solid line is $1.8\times x^{-1.2}$

\newpage

% GNUPLOT: LaTeX picture FIG-7: p(d\sqrt(t)) vs d\sqrt(t) for both in 3D
\setlength{\unitlength}{0.240900pt}
\ifx\plotpoint\undefined\newsavebox{\plotpoint}\fi
\sbox{\plotpoint}{\rule[-0.200pt]{0.400pt}{0.400pt}}%


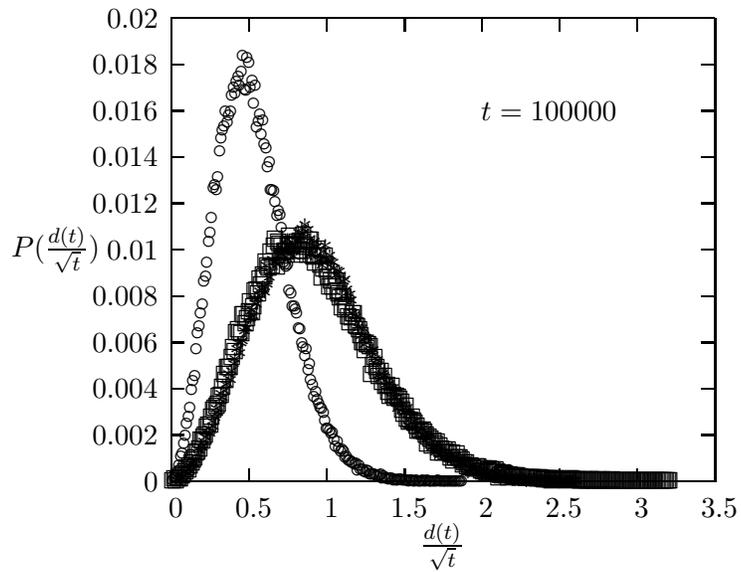
\noindent {\bf Fig-7.} The probability distribution of rescaled displacement 
in 3D continuum and in regular lattice after time, $t=100000$.
 For 3D continuum with  uniform length-step ($\Box$),  with random 
length-step ($\circ$), and for 3D regular lattice ($\ast$). 
\newpage

% GNUPLOT: LaTeX picture FIG-8 Prob vs rlim in 3D cont
\setlength{\unitlength}{0.240900pt}
\ifx\plotpoint\undefined\newsavebox{\plotpoint}\fi
\sbox{\plotpoint}{\rule[-0.200pt]{0.400pt}{0.400pt}}%
%

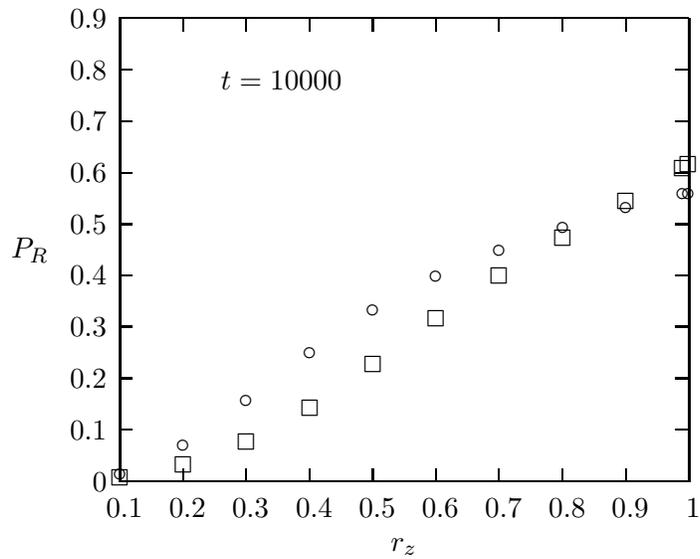
\noindent {\bf Fig-8.} The variation of probability of return for the first
 time  to the central zone with  radius of the zone, $r_z=.5$ in 3D continuum 
for time, $t=10000$.  For uniform length-step ($\Box$) and for random length-step ($\circ$).

\newpage

% GNUPLOT: LaTeX picture FIG-9
\setlength{\unitlength}{0.240900pt}
\ifx\plotpoint\undefined\newsavebox{\plotpoint}\fi
\sbox{\plotpoint}{\rule[-0.200pt]{0.400pt}{0.400pt}}%
%

\noindent {\bf Fig-9.} Log-log plot of the  probability distribution  of first 
return  time  to the central zone for a particular  radius of the zone, $r_z=.5$
 in 3D continuum  and for regular lattice for time, $t=10000$ and $N_s=100000$.
For uniform length-step ($\Box$), random length-step ($\circ$) and for regular
3D lattice ($\ast$). The solid line is $2.4\times x^{-1.32}$

\newpage

\newpage

\begin{center}{\bf References}\end{center}
\begin{enumerate}

\bibitem{binder} D. P. Landau and K. Binder, in {\it A guide to Monte
Carlo simulations in Statistical Physics}, Cambridge University Press,
(2005) pp 61

\bibitem{simple} A. Einstein, Ann Phys (Leipzeg)17, 549, (1990),
 S. Havlin and Ben-Avaham, Adv Phys 36, 695 (1987),
J. P. Bouchaud and A. Georges, Phys Rep, 195, 127 (1990).

\bibitem{poly} K. Binder, Avd. Polymer Sc. {\bf 112} (1994) 181

\bibitem{mono} K. Ohno and K. Binder, J. Stat. Phys. {\bf 64} (1991) 781

\bibitem{bethe} A. Giacometti, J. Phys: Math. Gen. {\bf 28} (1995) L13

\bibitem{bias} L. Scarpalezos, A. Kittas, P. Argyrakis, R. Cohen and
S. Havlin, Phys. Rev. E. {\bf 88} (2013) 012817

\bibitem{Denis} Denis Boyer and Citlali Solis-Salas, Phys. Rev. Lett. {\bf 112} (2014) 240601.

\end{enumerate}
\end{document}